\documentstyle[12pt]{article}

\begin{document}
%
\thispagestyle{empty}
\begin{flushright}
 UCT-TP 229/96 \\
hep-th/9607077
\end{flushright}

\vspace*{2cm}

\begin{center}
{\LARGE\bf Gravitational Phase Transition\\[1ex]
of Heavy Neutrino Matter}\\[1cm]
{\large Neven Bili\'{c} \footnote{On leave of absence from the
 Rudjer Bo\v{s}kovi\'{c} Institute, Zagreb, Croatia}\\[1ex]
and\\[1ex]
Raoul D.~Viollier \\[1ex]
Institute of Theoretical Physics and Astrophysics\\[1ex]
University of Cape Town, Rondebosch 7700, South Africa}\\

\vspace{3cm}

{\Large\bf Abstract}\\
\end{center}

\vspace{3ex}

\noindent

We study
the phase transition
of a system of self-gravitating neutrinos in the presence
of a large radiation density background
in the framework
of the Thomas-Fermi model.
We show  that, by cooling a non-degenerate gas of massive
neutrinos below some critical temperature, a condensed phase
emerges, consisting of quasi-degenerate supermassive neutrino
stars. These compact dark objects
could play an important role in structure formation in this
universe, as they might in fact provide the seeds for galactic
nuclei and quasi-stellar objects.

\newpage

A gas of massive fermions which interact only gravitationally
has interesting thermal properties which may have important
consequences for the early universe.
     The canonical and grand canonical ensembles
     for such a system
  have been shown to
have a nontrivial thermodynamical limit~\cite{Thir,Hert}.
Under certain conditions these systems
 will undergo a phase transition
 that is accompanied by gravitational collapse~\cite{Mess}.
 This phase transition
occurs uniquely for the attractive gravitational interaction of
neutral fermions.  There is indeed no such phase transition in
the case of charged fermions~\cite{Feyn}.  Of course, gravitational
condensation will also take place if the fermions have an
additional weak interaction, as neutrinos, neutralinos,
axions and other
weakly interacting massive particles generally do.
 Moreover, it is important to
note that the phase transition will occur, irrespective of the
magnitude of the initial density fluctuations in the fermion gas,
and, as we shall shortly demonstrate, irrespective of the
amount of background radiation.
 To be specific, we
henceforth assume that this  neutral fermion  is the heaviest
neutrino $\nu_{\tau}$, although this is not essential for most of the
subsequent discussion.

The ground state of a gravitationally condensed neutrino cloud,
with mass below the Chandrasekhar limit, is a cold neutrino
star~\cite{Luri,RDV6,RDV7,RDV8}, in which the degeneracy pressure
balances the gravitational attraction of the neutrinos.
Degenerate stars of neutrinos in the mass range between $m_{\nu}
= 10$ and  25 keV are particularly
interesting~\cite{RDV8}, as they could explain, without resorting
to the black hole hypothesis, at least some of the features that
are observed around supermassive compact dark objects, which are
reported to exist at the centers of a number of
galaxies~\cite{Tonr,Dres10,Dres11,Korm12,Korm13,Korm14} including
our own~\cite{Lacy15,Lacy16} and quasi-stellar objects (QSO)
\cite{Bell,Zeld,Blan,Bege}.  Indeed, there is little difference
between a supermassive black hole and a neutrino star of the same
mass near the Chandrasekhar limit, a few Schwarzschild radii away
from the object.

The existence of a quasi-stable neutrino in this mass range is
neither ruled out by particle and nuclear physics experiments nor by
direct astrophysical observations~\cite{RDV8}.  In the
early universe, however, it would lead to an early neutrino matter
dominated phase some time after nucleosynthesis and prior to
recombination.  In such a universe, the microwave background
temperature would be reached much too early to accommodate the
oldest stars in globular clusters, cosmochronology and the
Hubble expansion age, if the Standard Model of Cosmology is
correct. However, the early universe might have evolved quite
differently in the presence of such a heavy neutrino.  In
particular, it is conceivable that primordial neutrino stars have
been formed in local condensation processes during a
gravitational phase transition that must have occurred some time
between nucleosynthesis and recombination.  Aside from
reheating the gaseous phase of heavy neutrinos,
the latent heat
produced by the
condensed phase might have contributed partly to reheating the
radiation as well.  Moreover, the bulk part of the heavy neutrinos
(and antineutrinos) will have annihilated efficiently into light
neutrinos via the $Z^{0}$ in the interior of these supermassive
neutrino
stars~\cite{Luri,RDV6,RDV7,RDV8}.
 Since both these processes will
increase the age of the universe, or the time when the universe
reaches today's microwave background temperature~\cite{RDV8},
 it does not seem
excluded that a quasi-stable massive neutrino in the mass range
between  10  and 25
keV is compatible with the cosmological
observations~\cite{Luri,RDV8}.

\newpage

The purpose of
this paper is to study the formation of such a neutrino star
during a gravitational phase transition in an expanding
 universe at the time when
 the energy densities of neutrino matter and radiation
 are of comparable magnitude.


  We assume here that the
equilibrium distribution of the $\nu_{\tau}$ gas
 is spherically symmetric and
the energy density $\rho_{\gamma}$ of the radiation background
homogeneous.
At this stage
$\rho_{\gamma}$ consists of
photons and the two remaining relativistic
neutrino species
$\nu_{\mu}$ and
$\nu_e$,   is given by
\begin{equation}\label{eq01}
\rho_{\gamma} = \frac{a}{2} g_2 T^4_{\gamma},
\end{equation}
with
\begin{equation}\label{eq02}
g_N=2+\frac{7}{4}\left(\frac{4}{11}\right)^{4/3} N .
\end{equation}
The gravitational potential $V(r)$ satisfies the Poisson equation
\begin{equation}\label{eq00}
\Delta V = 4\pi G m_{\nu} (m_{\nu}n_{\nu}+\rho_{\gamma}),
\end{equation}
where the number density of $\tau$
neutrinos (including antineutrinos) of mass $m_{\nu}$ can be expressed
in terms of the Fermi-Dirac distribution at a finite temperature
$T$ as
\begin{eqnarray}\label{eq10}
n_{\nu}(r) & = & \frac{g_{\nu}}{4 \pi^{2}} \left(
2m_{\nu} T \right)^{3/2} I_{\frac{1}{2}} \left( \frac{\mu - V(r)}{T}
\right),
\end{eqnarray}
 with
\begin{eqnarray}\label{eq20}
I_{n} (\eta) & = & \int^{\infty}_{0} \frac{\xi^{n} d \xi}{1 +
e^{\xi - \eta}}.
\end{eqnarray}
$g_{\nu}$ denotes the combined spin degeneracy factors of
neutrinos and antineutrinos (i.e.\ $g_{\nu}$ is 2 or 4 for Majorana
 or Dirac neutrinos respectively), and $\mu$ is the chemical
potential.
It is convenient to
introduce the normalized reduced potential
\begin{equation}\label{21}
v = \frac{r}{m_{\nu}GM_{\odot}}(\mu-V),
\end{equation}
 $M_{\odot}$
being the solar mass,
and dimensionless variable $x=r/R_0$
with the scale factor
\begin{equation}\label{eq40}
R_0 =  \left( \frac{3 \pi } {4 \sqrt{2} m_{\nu}^{4}
g_{\nu} G^{3/2} M_{\odot}^{1/2}} \right)^{2/3} =
2.1377\;\;{\rm lyr} \left( \frac{17.2\;\;{\rm keV}}{m_{\nu}}
\right)^{8/3} g_{\nu}^{- 2/3}.
\end{equation}

Eq.(\ref{eq00}) then takes the simple form
\begin{equation}\label{eq50}
\frac{1}{x}
\frac{d^{2} v}{dx^{2}} = - \frac{3}{2} \beta^{-
3/2} I_{\frac{1}{2}} \left( \beta \frac{v}{x} \right)
-4\pi\frac{R_0^3\rho_{\gamma}}{M_{\odot}},
\end{equation}
where we have introduced the
normalized inverse temperature
defined as
\begin{equation}\label{51}
  \beta = T_{0}/T  ;\;\;\;\;\; T_0=m_{\nu} GM_{\odot} / R_0 .
\end{equation}
 At zero temperature
 we recover from eq.(\ref{eq50}) the
well-known Lan\'{e}-Emden differential equation [6,8]
\begin{eqnarray}\label{eq60}
\frac{d^{2} v}{dx^{2}} & = & - \frac{v^{3/2}}{\sqrt{x}}.
\end{eqnarray}

The solution of the differential equation (\ref{eq50}) requires
boundary conditions. We assume here that
 the neutrino gas is enclosed in a spherical
cavity of radius $R$ corresponding to $x_1=R/R_0$.
 We further require the total
neutrino mass to be $M_{\nu}$, and the total radiation mass
within the cavity
 to be $M_{\gamma}$, and we allow for the possibility
of a pointlike mass $M_{C}$ at the origin, which could be e.g.\ a
compact seed of other exotic matter.  $v(x)$ and
$v(x)$ is then related to
its derivative
at $x = x_1$ by
\begin{eqnarray}
v' (x_1)      & = & \frac{1}{x_1} \left( v (x_1) -
\frac{M_C+M_{\gamma} + M_{\nu}}{M_{\odot}} \right),
\label{eq100}
\end{eqnarray}
which in turn is related to
the chemical potential by $\mu = T_{0} v' (x_1)$.
$v(x)$ at $x=0$ is related to
 the point mass at the origin by
 $M_{C}/M_{\odot} = v(0)$.

Similar to the case of the Lan\'{e}-Emden equation,
it is easy to show that eq.(\ref{eq50}) has a scaling property:
 if $v(x)$
is a solution of eq.(\ref{eq50}) at a temperature $T$
and a cavity radius $R$, then
$\tilde{v} (x) = A^{3} v (Ax)$ with $(A > 0)$ is also a solution
at the temperatures $\tilde{T} = A^{4} T$,
$\tilde{T}_{\gamma} = A^{4} T_{\gamma}$
and the cavity radius $\tilde{R}=R/A$.

It is important to note that only those
solutions that
 minimize  the free energy  are physical.
 The free energy functional is defined as~\cite{Hert}
\begin{eqnarray}
F[n] & = & \mu[n] N_{\nu}-W[n]
\nonumber  \\
     & - &
 Tg_{\nu}\int\frac{d^3rd^3p}{(2\pi)^3}
\ln( 1+\exp \left(-\frac{p^2}{2m_{\nu}T}-
\frac{V[n]}{T}+\frac{\mu[n]}{T}\right)),
\label{eq120}
\end{eqnarray}
where
\begin{equation}\label{eq130}
V[n] =  -Gm_{\nu}
\int d^3r'\frac{m_{\nu}n(r')+\rho_{\gamma}}{|\bf{r}-\bf{r}'|},
\end{equation}
and
\begin{equation}\label{eq131}
W[n] =  -\frac{1}{2}Gm_{\nu}^2
\int d^3rd^3r'\frac{n(r)n(r')}{|\bf{r}-\bf{r}'|}.
\end{equation}
The chemical potential in eq. (\ref{eq120}) varies with
density so that
 the number of neutrinos $N_{\nu}=M_{\nu}/m_{\nu}$
is kept fixed.

All the relevant thermodynamical quantities
such as
number density, pressure,
free energy, energy and entropy
can be expressed in terms of $v/x$
\begin{eqnarray}
n_{\nu}(x)
      \!&\!=\!&\!
        \frac{M_{\odot}}{m_{\nu} R_0^3}
                     \frac{3}{8\pi}
                     \beta^{-3/2} I_{\frac{1}{2}}
                     \left( \beta \frac{v}{x} \right),
 \label{eq140}\\[.2cm]
P_{\nu}(x)
      \!&\!=\!&\!
           \frac{M_{\odot} T_0}{m_{\nu} R_0^3}
                     \frac{3}{8\pi}
                     \beta^{-5/2} I_{\frac{3}{2}}
                     \left( \beta \frac{v}{x} \right) =
                     \frac{2}{3} \varepsilon_{\rm kin} (x),
\label{eq150}
\end{eqnarray}
\newpage
\begin{eqnarray}
F
      \!&\!=\!&\!
\frac{1}{2}\mu(N_{\nu}+N_{\gamma})
 +\frac{3}{5}M_{\gamma}^2\frac{1}{R}
\nonumber  \\
      \!&\!+\!&\!
 \frac{1}{2}T_0R_0^3\int d^3x(n_{\nu}-n_{\gamma})\frac{v(x)-v(0)}{x}
      -R_0^3\int d^3x P_{\nu}(x),
\label{eq160}\\[.2cm]
E
      \!&\!=\!&\!
\frac{1}{2}\mu(N_{\nu}+N_{\gamma})
 +\frac{3}{5}M_{\gamma}^2\frac{1}{R}
\nonumber  \\
      \!&\!-\!&\!
 \frac{1}{2}T_0R_0^3\int d^3x\left[(n_{\nu}+n_{\gamma})\frac{v(x)}{x}
 +(n_{\nu}-n_{\gamma})\frac{v(0)}{x}\right]
      +R_0^3\int d^3x
 \varepsilon_{\rm kin}(x),
\label{eq170}\\[.2cm]
 S
      \!&\!=\!&\!
     \frac{1}{T}(E-F).
\label{eq180}
\end{eqnarray}

In eqs. (\ref{eq160}) and (\ref{eq170}) we have introduced
the effective radiation number
$N_{\gamma}=M_{\gamma}/m_{\nu}$,
and the effective radiation number density
$n_{\gamma}=\rho_{\gamma}/m_{\nu}$.

%
%

We now turn to the numerical study of a
  system
of self-gravitating massive neutrinos
with arbitrarily chosen
total mass $M=10 M_{\odot}$ varying the cavity radius $R$
and the radiation mass $M_{\gamma}$.
Owing to the scaling
properties, the system may be rescaled to any physically
interesting mass.
For definiteness, the $\nu_{\tau}$ mass
is chosen as
 $m_{\nu}=17.2$ keV
 which is about the central value of the mass
 region between 10 and 25 keV \cite{RDV8}
 that is interesting for our scenario.

In fig. 1 we present our results for a
gas of neutrinos
 in a cavity of
radius $R=100 R_0$,
and with no
background radiation
i. e. with $M_{\gamma}=0$.
We find the
three distinct solutions
in the temperature interval
 $T=(0.049-0.311)T_0$
 of which only two are physical,
 namely those
 for which the free-energy
  assumes a minimum.
 The density distributions,
  corresponding to
 such two solutions are shown in the first plot
 in fig. 1.
 We refer to the solution
 that exists above the
 mentioned interval at arbitrary high temperature
 as ``gas", while the solution which
  exists at  low
 temperatures and eventually becomes a degenerate
 Fermi gas at $T=0$, we refer to as ``condensate".

 In fig. 1 we also plot
various extensive
thermodynamical quantities
(per neutrino)
as functions of neutrino temperature.
  The phase transition takes place at the point
 $T_t$ where the free energy of the gas and
 condensate become equal.
The transition temperature
 $T_t=0.19442T_0$ is
indicated by the dotted line
 in the free energy plot.
 The top dashed curve in the same plot
   corresponds to the unphysical solution.
 At $T=T_t$ the energy and the entropy have a discontinuity.

Next we study a system of massive neutrino gas in
a radiation background.
In the course of universe expansion,
the heavy neutrino becomes nonrelativistic
at a time $t_{NR}$ corresponding to the temperature
$T_{NR}=m_{\nu}$. At that time the radiation dominates
the matter by about a factor of 15.
A neutrino cloud with the total mass e.g.
$10^9 M_{\odot}$,
which is by about factor of 10 below the
Chandrasekhar limit for
 $m_{\nu}=17.2$ keV and
 $g_{\nu}=4$,
 would have a radius
\begin{equation}
R_{NR}=2[G(M_{\nu}+M_{\gamma})t_{NR}^2]^{1/3},
\label{310}
\end{equation}
yielding $R_{NR}=0.265$ light days.
As we shall shortly see  this
is way below the critical value and therefore
there exists
only one solution
for any fixed temperature.
As the universe expands and cools down
the relative amount of radiation mass
decreases as
\begin{equation}
M_{\gamma}=15M_{\nu}\frac{R_{NR}}{R}.
\label{320}
\end{equation}
Concerning the $R$ dependence of the temperature
we assume that during the
 radiation domination the temperature $T$ of the
neutrino gas is fixed by the radiation heat bath
so that it continues to decrease  as $1/R$ until
 the neutrino matter starts to dominate.
At that point the system being in gaseous
phase will have the entropy per particle
$S/N_{\nu}=7.60$.
From then on,
the $T$  will not be coupled to the radiation
heat bath anymore and will decrease
with $R$ according to the adiabatic expansion
of the neutrino matter itself.

Shortly after the matter starts to dominate,
we reach the region of instability and
the first order phase transition takes place.
Fig 2. shows how the entropy behaves around the critical
region.
We find as we decrease the radius
that the first order phase transition becomes weaker
 and eventually disappears at $R=R_c=5.34$ light days
 corresponding to the critical temperature
$T_c=0.0043m_{\nu}$ and the radiation mass
relative to the neutrino mass
$M_{\gamma}=0.67M_{\nu}$.
This is the critical point of a second order
phase transition. This behavior is typical for
a mean-field type of models \cite{bil}.

The initial entropy of 7.6 per neutrino
drops to  0.75 so that the latent heat per particle
yields
$\Delta E= T \Delta S= 0.016m_{\nu}$.
Thus,
the condensate formation is accompanied by
a release of considerable amount of
energy which will reheat
the radiation environment.

\vspace{0.2in}
{\large \bf Acknowledgement}

We acknowledge useful discussions with
 D. Tsiklauri and G.J. Kyrchev.
\vspace{0.4in}

\vspace{0.5in}
\noindent {\bf Figure Captions:}
\begin{description}
\medskip
\item{1:}
Density distribution normalized to unity
  for  condensate-like
(solid line) and  gas-like (dashed line) solutions
at $T=T_t$.
Free energy, energy
and entropy per particle as a function of temperature.
Temperature, energy and
free energy, are
in units of $T_0$.
\vskip 3pt
\item{2:}
Entropy per particle as a function of temperature for various
radii (in light days) of the neutrino cloud with
$M_{\nu}=10^9M_{\odot}$.
\end{description}
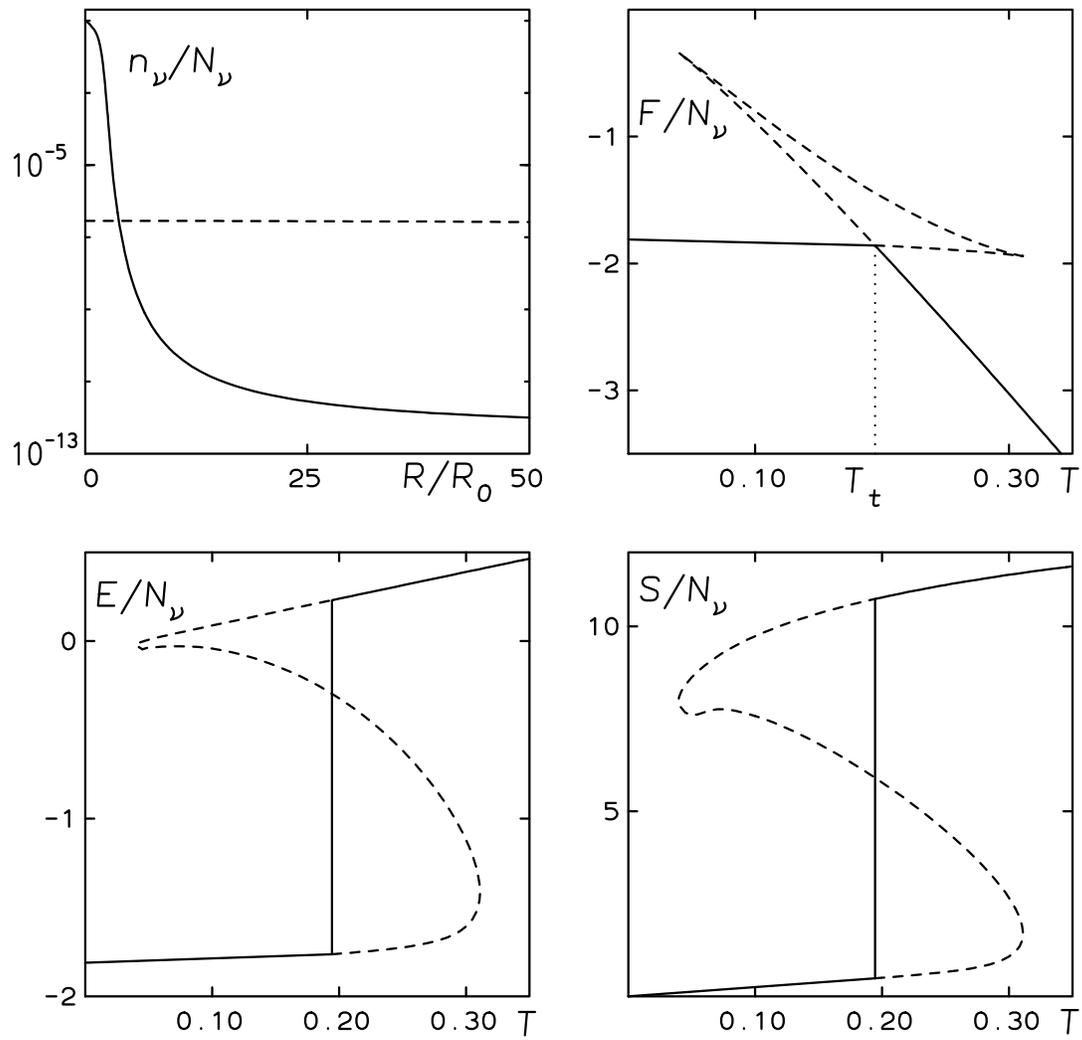
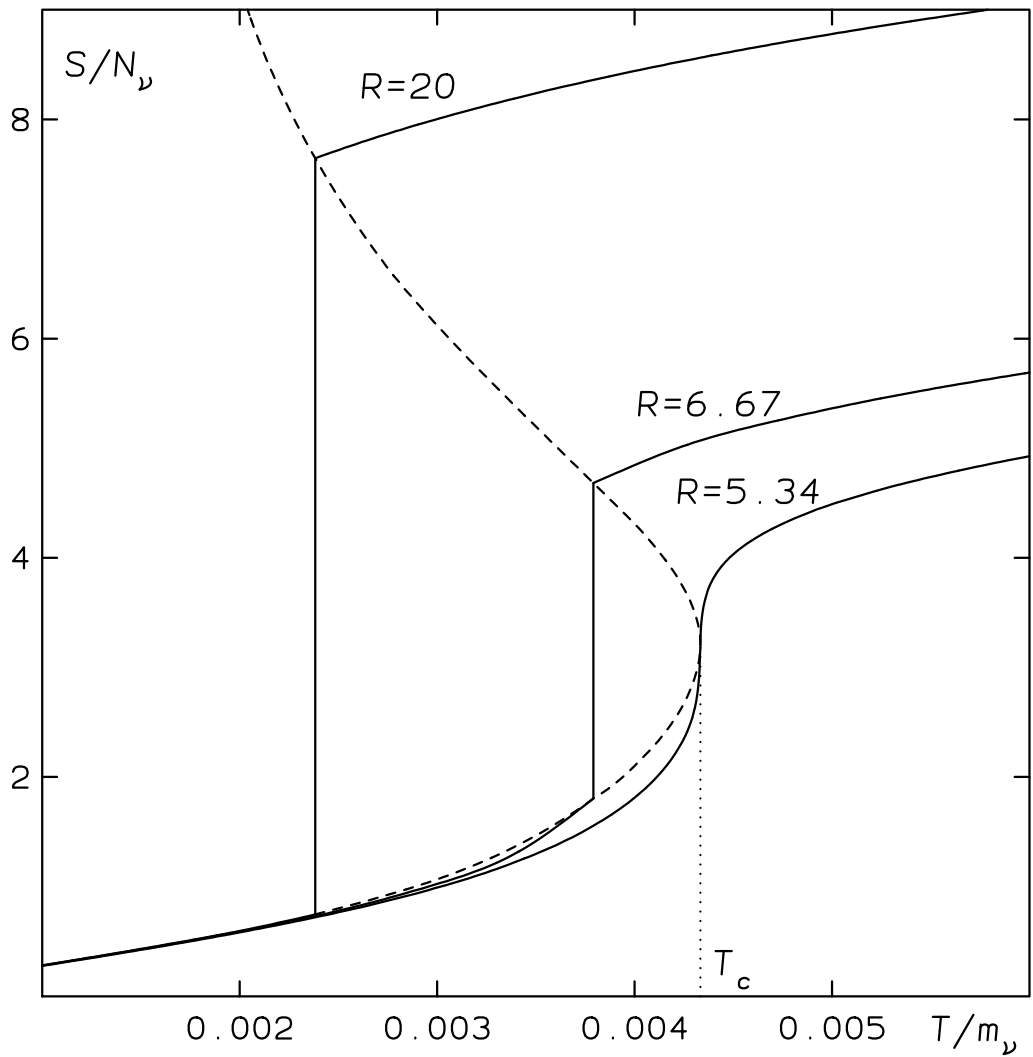
\end{document}